\documentclass[aps,showpacs,preprintnumbers,amsmath,amssymb,nofootinbib]{revtex4}
\usepackage{bm}

\begin{document}

\title{Gravitational waves generated during inflation from a 5D vacuum theory of gravity in a de Sitter expansion}

\author{Jos\'e Edgar Madriz Aguilar\footnote{E-mail address: jemadriz@fisica.ufpb.br}}
\affiliation{Departamento de F\'{\i}sica, Universidade Federal da
Para\'{\i}ba. C.P. 5008-CEP: 58059-970, \\ Jo\~{a}o Pessoa, PB
58059-970 Brazil.}

\vskip .5cm
\begin{abstract}
In this letter we study the generation of gravitational waves during inflation from a 5D vacuum theory of gravity. Within this 
formalism, on an effective 4D de Sitter background, we recover the typical results obtained with 4D inflationary theory in general 
relativity, for the amplitude of gravitational waves generated during inflation. We also obtain a range of values for the 
amplitude of tensor to scalar ratio which is in agreement with COBE observations.
\end{abstract}


\maketitle

\section{Introduction}
It is well known from the literature and some review papers that inflation predicts the existence of gravitational waves generated 
through a similar mechanism as the one that generates  energy density perturbations \cite{WW1}. These cosmological gravitational 
waves give valuable information about the very early universe. Gravitons can propagate through the expanding universe, living 
prints on the power spectrum and polarization of the cosmic microwave background radiation (CMBR).
During the last years there has been significant research on gravitational waves in both theoretical and detection aspects, 
providing new advances in the searching of gravitational waves produced from astrophysical and cosmological sources. Particularly, 
the detection of inflationary gravitational waves is one of the challenges of LISA \cite{WW2} and other experimental projects 
\cite{WW3}. In addition, the interest in the study of gravitational waves during inflation from a higher dimensional theories has 
been increasing. \\

Inflationary cosmology has been approached in different higher dimensional frameworks, among them some of the most relevant are  
the braneworlds \cite{au} and the induced matter theory \cite{WW4}. According to braneworld cosmology the universe is considered 
as a brane embedded in a higher dimensional space-time called the bulk. In this approach only gravity and other exotic matter as 
some scalar fields (like dilaton) can propagate through the bulk, while matter is confined on the brane \cite{bra1}. Many papers 
studying gravitational waves generated during inflation from braneworlds have been published \cite{WW5}. An extensive analysis of 
gravitational waves into a period of quasi-de Sitter expansion has been made in \cite{WW6}.
On the other hand, the induced matter theory is a kind of noncompact Kaluza-Klein theories in 5D, since the fifth dimension is 
considered extended. In the 90's Paul Wesson, J. Ponce de Leon and collaborators, based on the Campbell Magaard theorem 
\cite{campbell,magaard}, proposed that matter and its dynamics in 4D, can be geometrically induced  from a 5D apparent vacuum 
defined by $R_{AB}=0$, which automatically implies $G_{AB}=0$ \cite{lwess}. In our notation conventions henceforth capital Latin 
indices $A,B,...=$ run from $0$ to $4$, whereas small Latin indices $i,j,...$, from $1$ to $3$.
In the context of this theory it has been shown that gravitational waves in 5D can be approached as in 4D, by using a linearized 
metric and the harmonic gauge condition \cite{WW7}. With the same spirit, using ideas of the induced matter theory a novel 
formalism for describing inflation from a 5D apparent vacuum has been recently introduced \cite{WW8}. Within this formalism the 4D 
scalar inflationary potential related to the inflaton field is geometrically generated from a 5D apparent vacuum. The interesting 
here is that the 4D effective induced inflationary potential depends only of derivatives of the inflaton field with respect to the 
fifth coordinate, having this way a pure geometrical nature. In the case of a de Sitter expansion, the existence of a non-constant 
inflationary potential is possible in this formalism \cite{WW8}.  Gauge invariant metric fluctuations in a de Sitter expansion, 
also within this formalism, has been studied in \cite{WW9}. One of the most important results obtained in \cite{WW9} is that in 
this formalism the small deviation from the scale invariance of the spectrum is due to the existence of the fifth dimension. With 
the aim of studying the complementary part of the analysis made in \cite{WW9}, in this letter we shall study the generation of 4D 
gravitational waves during inflation in a de Sitter expansion, induced from a 5D space-time in apparent vacuum. \\

\section{5D tensor perturbations}

In order to describe gravitational waves from a 5D apparent vacuum we consider the background metric \cite{PLB}
\begin{equation}\label{ww1}
dS_{b}^{2}=\psi^{2}dN^{2}-\psi^{2}e^{2N}dr^{2}-d\psi^{2},
\end{equation}
where $dr^{2}=E_{ij}dX^{i}dX^{j}$ being $E_{ij}$ the Euclidean metric and $(X^{i})$ dimensionless spatial coordinates. Besides, 
the time like coordinate $N$ is dimensionless and the space like fifth coordinate $\psi$ has spatial units. This metric has the 
property to be Riemann flat i.e. it satisfies $R^{A}\,_{BCD}=0$.

We consider a perturbed 5D line element of the form
\begin{equation}\label{tp1}
dS^{2}=\psi^{2}dN^{2}-\psi^{2}e^{2N}\left(\delta _{ij}+ \Pi _{ij}\right)dx^{i}dx^{j}-d\psi^{2},
\end{equation}
where $\Pi _{ij}$ is a transverse traceless tensor and therefore it satisfies $tr(\Pi _{ij})=\Pi^{i}\,_{i}=0$ and  $\partial _{i} 
\Pi^{ij}=0$. Thus, we can write the spatial components of the metric as $g_{ij}=-\psi^{2}e^{2N}\left(\delta _{ij}+\Pi 
_{ij}\right)$, so that we can consider the linear approximation of the contravariant metric $g^{ij}\simeq -\psi^{-2}e^{-2N}(\delta 
^{ij}-\Pi^{ij})$. \\

The equation of motion for the tensor modes is obtained by using the linearized 5D Einstein equations in vacuum $\delta R_{AB}=0$, 
and for the case of the tensor modes associated with (\ref{tp1}), these reduce simply to
\begin{equation}\label{tp2}
\delta R_{ij}=0.
\end{equation}
Hence, the equation of motion for the 5D tensor modes is
\begin{equation}\label{tp3}
\Pi^{i}\,_{j,NN}+3\Pi^{i}\,_{j,N}+e^{-2N}\left(\Pi^{i}\,_{k,j}\,^{k}+\Pi 
_{jk}\,^{,ik}-\Pi^{i}\,_{j,k}\,^{k}\right)-4\psi\Pi^{i}\,_{j,\psi}-\psi^{2}\Pi^{i}\,_{j,\psi\psi}=0.
\end{equation}
Expressing this equation in terms of the Fourier modes
\begin{equation}\label{tp5}
\Pi^{i}\,_{j}(N,\vec{r},\psi)=\frac{1}{(2\pi)^{3/2}}\int d^{3}k_{r}\int dk_{\psi}\sum _{\lambda 
}\,_{(\lambda)}e^{i}\,_{j}\left[a_{k_{r}k_{\psi}}^{(\lambda)}e^{i\vec{k}_{r}\cdot\vec{r}}\zeta _{k_{r}k_{\psi}}(N,\psi)+a_{k_{r}k
_{\psi}}^{(\lambda)\,\dagger}e^{-i\vec{k}_{r}\cdot\vec{r}}\zeta _{k_{r}k_{\psi}}^{*}(N,\psi)\right]
\end{equation}
with the asterisk $*$ denoting complex conjugate, $\lambda$ counting the number of degrees of freedom, and 
$a_{k_{r}k_{\psi}}^{(\lambda)\,\dagger}$ and $a_{k_{r}k_{\psi}}^{(\lambda)}$ being the creation and annihilation operators 
respectively, which satisfy the algebra
\begin{eqnarray}\label{tp6}
\left[a_{k_{r}k_{\psi}}^{(\lambda)},a_{k'_{r}k'_{\psi }}^{(\lambda ')\,\dagger}\right]&=&g^{\lambda\lambda '}\delta 
^{(3)}\left(\vec{k}_{r}-\vec{k}_{r}'\right)\delta(\psi - \psi ')\\
\label{tp7a} \left[a_{k_{r}k_{\psi}}^{(\lambda)},a_{k'_{r}k'_{\psi}}^{(\lambda 
')}\right]&=&\left[a_{k_{r}k_{\psi}}^{(\lambda)\,\dagger},a_{k'_{r}k'_{\psi}}^{(\lambda ')\,\dagger}\right]=0
\end{eqnarray}
and where $_{(\lambda)}e^{ij}$ is the polarization tensor that satisfies
\begin{equation}\label{tp7}
^{(\lambda)}e_{ij}=\,^{(\lambda)}e_{ji},\qquad k^{i}\,^{(\lambda)}e_{ij}=0,\qquad ^{(\lambda)}e_{ii}=0,\qquad 
^{(\lambda)}e_{ij}(-\vec{k}_{r})=\,^{(\lambda)}e_{ij}^{*}(\vec{k}_r).
\end{equation}
By inserting (\ref{tp5}) in the equation of motion (\ref{tp3}) we obtain
\begin{equation}\label{tp8}
\stackrel{\star\star}{\zeta}_{k_{r}k_{\psi}}+3\stackrel{\star}{\zeta}_{k_{r}k_{\psi}}+e^{-2N}k_{r}^{2}\zeta 
_{k_{r}k_{\psi}}-\left[4\psi\frac{\partial}{\partial \psi}+\psi^{2}\frac{\partial ^{2}}{\partial \psi^{2}}\right]\zeta 
_{k_{r}k_{\psi}}=0
\end{equation}
which is the equation of motion for the 5D tensor modes $\zeta _{k_{r}k_{\psi}}(N,\psi)$ and where $\star$ denotes derivative with 
respect to the time like coordinate $N$. Decomposing the tensor modes $\zeta_{k_{r}k_{\psi}}(N,\psi)$ into Kaluza-Klein modes
\begin{equation}\label{tp10}
\zeta _{k_{r}k_{\psi}}(N,\psi)=\int dm\, \zeta _{m}(N)\xi _{m}(\psi),
\end{equation}
equation (\ref{tp8}) is replaced by the system
\begin{eqnarray}\label{ntp1}
\stackrel{\star\star}{\zeta}_{m}+3\stackrel{\star}{\zeta}_{m}+\left(k_{r}^{2}e^{-2N}+m^{2}\right)\zeta _{m}&=&0\\
\label{ntp2}
\psi^{2}\frac{\partial^{2}\xi _{m}}{\partial \psi^2}+4\psi\frac{\partial \xi _{m}}{\partial \psi}+m^{2}\xi _{m}&=&0
\end{eqnarray}
where for simplicity we have suppressed the sub-index $k_{r}k_{\psi}$. The parameter $m^{2}$ corresponds to the square of the 
KK-mass as measured by a 5D observer. In order to simplify the structure of (\ref{ntp1}) we consider the field transformation 
$\zeta _{m}(N,\vec{r},\psi)=e^{-3N/2}\chi _{m}(N)$. Hence the dynamical equation (\ref{ntp1}) yields
\begin{equation}\label{tp9}
\stackrel{\star\star}{\chi}_{m}+\left[k_{r}^{2}e^{-2N}+\left(m^{2}-\frac{9}{4}\right)\right]\chi _{m}=0,
\end{equation}
whose general solution is given by
\begin{equation}\label{tp13}
\chi _{m}(N)=A_{1}{\cal H}_{\nu}^{(1)}[x(N)]+A_{2}{\cal H}_{\nu}^{(2)}[x(N)],
\end{equation}
being $\nu =(1/2)\sqrt{9-4m^{2}}$ and $x(N)=k_{r}e^{-N}$. The functions ${\cal H}_{\nu}^{(1)}[x(N)]$ and ${\cal 
H}_{\nu}^{(2)}[x(N)]$ denote the first and second kind Hankel functions respectively, and $A_{1}$, $A_{2}$ are integration 
constants. Analogously, the dynamical equation that depends of the fifth coordinate (\ref{ntp2}), by using the transformation
$\xi_{m}(\psi)=\left(\psi _{0}/\psi\right)^{2}P_{m}(\psi)$, acquires the form
\begin{equation}\label{ntp4}
\frac{\partial ^{2}P_{m}}{\partial \psi^2}+\left(\frac{m^{2}-2}{\psi^2}\right)P_{m}=0,
\end{equation}
with the sub-index $0$ denoting value at the end of inflation. Solving (\ref{ntp4}) we obtain
\begin{equation}\label{ntp5}
P_{m}(\psi)=B_{1}\psi^{(1/2)\left(1+\sqrt{9-4m^{2}}\right)}+B_{2}\psi^{(1/2)\left(1-\sqrt{9-4m^2} \right)},
\end{equation}
where $B_{1}$ and $B_{2}$ are integration constants. To ensure that the algebra given by (\ref{tp6}) and (\ref{tp7a}) to be 
satisfied, we use the normalization condition for the modes $\zeta _{k_{r}k_{\psi}}(N,\psi)$
\begin{equation}\label{ntp6}
\zeta 
_{k_{r}k_{\psi}}\left(\stackrel{\star}{\zeta}_{k_{r}k_{\psi}}\right)^{*}-\left(\stackrel{\star}{\zeta}_{k_{r}k_{\psi}}\right)\zeta 
_{k_{r}k_{\psi}}^{*}=i.
\end{equation}
Since (\ref{tp10}) is valid, the condition (\ref{ntp6}) is equivalent to the requirements
\begin{eqnarray}\label{ntp7}
&&\zeta _{m}\stackrel{\star}{\zeta}_{m}^{*}-\stackrel{\star}{\zeta}_{m}\zeta _{m}^{*}=i/a_{0}^{3},\\
\label{ntp8}
&& 2\int _{0}^{\psi _0}\xi _{m}\xi _{m}^{*}d\psi =1.
\end{eqnarray}
Thus using the Bunch-Davies condition ($A_{1}=0$) and (\ref{ntp7}), the normalized solution of (\ref{tp9}) is
\begin{equation}\label{ntp8}
\chi _{m}(N)=i\sqrt{\frac{\pi}{4}}{\cal H}_{\nu}^{(2)}[k_{r}e^{-N}].
\end{equation}
In order to study consequences of (\ref{ntp5}) in the KK-modes, we re-express the equation (\ref{ntp2}) in terms of the conformal 
spatial fifth coordinate $z=\int d\psi/\psi$. Furthermore using the transformation $\xi _{m}(z)=e^{-(3/2)z}L_{m}(z)$, the 
expression (\ref{ntp2}) now becomes
\begin{equation}\label{ntp9}
\frac{d^{2}L_{m}}{dz^2}+\left(m^{2}-\frac{9}{4}\right)L_{m}=0.
\end{equation}
From this equation is easy to probe that for $m<3/2$ the KK-modes are unstable and divergent at infinity. For $m>3/2$ the KK-modes 
are coherent ($L_{m}\sim e^{i\sqrt{m^{2}-9/4}\,z}$), which is not surprising since the 5D background metric is Riemann flat. The 
zero-mode ($m=0$) is finite and thus normalizable through the condition (\ref{ntp8}). Note that this results are the same ones 
that those obtained within the braneworld formalism in \cite{WW6}. Hence, normalizing the zero-mode and using the Bunch-Davies 
condition ($B_{2}=0$), we obtain that $B_{1}=\sqrt{(1/2)}\,\psi _{0}^{-2}$.

\section{The effective 4D gravitational waves}
Now we are able to study the effective 4D induced dynamics for the tensor perturbations of the metric (\ref{tp1}), in an effective 
4D de Sitter background. As we will see along the present section, as it is typical in the induced matter theory, in the 4D part 
of the analysis the contribution of the fifth coordinate is no treated separately from the usual 4D part, as it is usually done in 
braneworld cosmology. Instead of that, in 4D the fifth coordinate is responsible of the appearance of new dynamical (and also 
geometrical) terms in the equation of motion for the effective 4D tensor metric fluctuations, all of them with time and spatial  
dependence. Therefore, it is possible to treat this terms together with the rest of them in the usual 4D form.
\subsection{ The effective 4D metric}
In order to describe the dynamics of the induced effective 4D tensor modes in physical coordinates, we use the coordinate 
transformation
\begin{equation}\label{4tp1}
N=Ht,\qquad r=HR,\qquad \psi =\psi
\end{equation}
where $t$ is the cosmic time, $R$ has spatial units and $H$ is the constant Hubble parameter. Hence the 5D background line element 
(\ref{ww1}) now has the form
\begin{equation}\label{4tp2}
dS_{b}^{2}=\left(H\psi\right)^{2}\left[dt^{2}-e^{2Ht}dR^{2}\right]-d\psi^2,
\end{equation}
which is effectively the Ponce de Leon Metric \cite{Pleon1} and describes a 5D generalized de Sitter expansion of the universe. In 
this new coordinates the perturbed line element (\ref{tp1}) is
\begin{equation}\label{4ntp1}
dS^{2}=(H\psi)^{2}\left[dt^{2}-e^{2Ht}(\delta _{ij}+\bar{\Pi}_{ij})dx^{i}dx^{j}\right] - d\psi ^{2},
\end{equation}
being $\bar{\Pi}_{ij}=\bar{\Pi}_{ij}(t,\vec{R},\psi)$. By taking the foliation $\psi =H^{-1}$, the line element (\ref{4tp2}) 
yields
\begin{equation}\label{4tp3}
ds_{b}^{2}=dt^{2}-e^{2Ht}dR^{2},
\end{equation}
which is an effective 4D metric that describes a de Sitter expansion of the universe.
In a similar manner taking the foliation in (\ref{4ntp1}), the effective 4D tensor-perturbed line element (\ref{tp1}) onto the 
hypersurfaces $\psi =H^{-1}$ becomes
\begin{equation}\label{4tp4}
ds^{2}=dt^{2}-e^{2Ht}(\delta _{ij}+h_{ij})dx^{i}dx^{j},
\end{equation}
where $h_{ij}(t,\vec{R})=\bar{\Pi}_{ij}(t,\vec{R},\psi)|_{\psi =H^{-1}}$ describes the tensor perturbations with respect to the 
effective 4D background metric (\ref{4tp3}).

\subsection{Dynamics of the effective 4D tensor modes in a de Sitter expansion}

In order to study the dynamics of the 4D tensor-fluctuations $h_{ij}(t,\vec{R})$ in a de Sitter expansion we apply the coordinate 
transformation (\ref{4tp1}) into the expression (\ref{tp3}), followed by taking the foliation $\psi =H^{-1}$. Thus, we obtain
\begin{equation}\label{4tp5}
\ddot{h}^{i}\,_{j}+3H\dot{h}^{i}\,_{j}+e^{-2Ht}\left(h^{i}\,_{k,j}\,^{k}+h_{jk}\,^{,ik}-h^{i}\,_{j,k}\,^{k}\right)-\left.\left[ 
\frac{4}{\psi}\bar{\Pi}^{i}\,_{j,\psi}+\bar{\Pi}^{i}\,_{j,\psi\psi}\right]\right|_{\psi =H^{-1}}=0,
\end{equation}
where $(\cdot)$ denotes derivative with respect to the cosmic time $t$. Note that the last term is an extra term that does not 
appears in the typical 4D treatments of gravitational waves during inflation, and that it has a geometric origin. Calculating the 
extra term as a function of $h^{i}\,_{j}(t,\vec{R})$, the equation (\ref{4tp5}) acquires the form
\begin{equation}\label{4tp6}
\ddot{h}^{i}\,_{j}+3H\dot{h}^{i}\,_{j}+e^{-2Ht}\left(h^{i}\,_{k,j}\,^{k}+h_{jk}\,^{,ik}-h^{i}\,_{j,k}\,^{k}\right)+m^{2} 
H^{2}h^{i}\,_{j}=0.
\end{equation}
Here $m^{2}$ corresponds to the squared of the KK-mass, as measured by an observer on the effective 4D de Sitter hypersurfaces 
$\psi= H^{-1}$. Equation (\ref{4tp6}) describes the dynamics of the effective 4D tensor modes $h^{i}\,_{j}(t,\vec{R})$. Now, for 
simplicity, we find convenient to work with the conformal time $\eta =-1/(aH)$, where $a(t)=e^{Ht}$ is the scale factor. Therefore 
the equation (\ref{4tp6}) gives
\begin{equation}\label{4tp7}
(h^{i}\,_{j})^{''}+2{\cal H}(\eta)(h^{i}\,_{j})^{'}+\left(h^{i}\,_{k,j}\,^{k}+h_{jk}\,^{,ik}-h^{i}\,_{j,k}\,^{k}\right)+m^{2} 
{\cal H}^{2}(\eta)h^{i}\,_{j}=0,
\end{equation}
where $(')$ denotes derivative with respect to the conformal time $\eta$ and ${\cal H}(\eta)=a'(\eta)/a(\eta)$. Implementing the 
field transformation $h^{i}\,_{j}(\eta ,\vec{R})=a^{-1}(\eta)W^{i}\,_{j}(\eta ,\vec{R})$ the equation (\ref{4tp7}) will be
\begin{equation}\label{4tp8}
(W^{i}\,_{j})^{''}+\left(W^{i}\,_{k,j}\,^{k} + W_{jk}\,^{,ik} - W^{i}\,_{j,k}\,^{k}\right)+\left[m^{2} {\cal 
H}^{2}(\eta)-\frac{a^{''}(\eta)}{a(\eta)}\right]W^{i}\,_{j}=0.
\end{equation}
We can express the tensor field $W^{i}\,_{j}(\eta,\vec{R})$ as a Fourier expansion
\begin{equation}\label{4tp9}
W^{i}\,_{j}(\eta,\vec{R})=\frac{1}{(2\pi)^{3/2}}\int d^{3}k_{R}\sum _{\alpha 
=+,\times}\,_{(\alpha)}e^{i}\,_{j}\left[b_{k_{R}}^{(\alpha)}e^{i\vec{k}_{R}\cdot\vec{R}}{\cal 
G}_{k_R}(\eta)+b_{k_R}^{(\alpha)\,\dagger}e^{-i\vec{k}_{R}\cdot\vec{R}}{\cal G}_{k_R}^{*}(\eta)\right]
\end{equation}
with the creation and annihilation operators $b_{k_R}^{(\alpha)}$ and $b_{k_R}^{(\alpha)\,\dagger}$ respectively, satisfying
\begin{equation}\label{4tp10}
\left[b_{k_R}^{(\alpha)},b_{k'_{R}}^{(\alpha ')\,\dagger}\right]=g^{\alpha\alpha '}\delta 
^{(3)}\left(\vec{k}_{R}-\vec{k}'_{R}\right),\qquad \left[b_{k_R}^{(\alpha)},b_{k'_{R}}^{(\alpha 
')}\right]=\left[b_{k_{R}}^{(\alpha)\,\dagger},b_{k'_{R}}^{(\alpha ')\,\dagger}\right]=0,
\end{equation}
and where the following properties for the polarization tensor $_{(\alpha)}e^{ij}$ are valid
\begin{equation}\label{4tp11}
^{(\alpha)}e_{ij}=\,^{(\alpha)}e_{ji},\quad k_{R}^{i}\,^{(\alpha)}e_{ij}=0,\quad ^{(\alpha)}e_{ii}=0,\quad 
^{(\alpha)}e_{ij}[-\vec{k}_{R}]=\,^{(\alpha)}e_{ij}^{*}[\vec{k}_{R}].
\end{equation}
Hence, the dynamical equation for the $k_{R}$-tensor modes ${\cal G}_{k_R}(\eta)$ is given by
\begin{equation}\label{4tp12}
{\cal G}^{''}_{k_R}+\left[k_{R}^{2}-\frac{(2-m^{2})}{\eta ^2}\right]{\cal G}_{k_R}=0,
\end{equation}
whose general solution is
\begin{equation}\label{4tp13}
{\cal G}_{k_R}(\eta)=\sqrt{-\eta}\left(F_{1}{\cal H}_{\nu_{T}}^{(1)}[-\eta k_R]+F_{2}{\cal H}^{(2)}_{\nu _T}[-\eta k_R]\right).
\end{equation}
where $F_1$ and $F_2$ are integration constants and $\nu_{T}=(1/2)\sqrt{9-4m^{2}}$. Here ${\cal H}_{\nu_T}^{(1)}$ and ${\cal 
H}_{\nu_T}^{(2)}$ denote the first kind Hankel functions. The normalization condition for the tensor modes ${\cal G}_{k_R}$ will 
be
\begin{equation}\label{4tp14}
{\cal G}_{k_{R}}\left({\cal G}'_{k_R}\right)^{*}-{\cal G}'_{k_R}{\cal G}_{k_R}^{*}=i.
\end{equation}
Under the Bunch-Davies vacuum condition, the normalized solution of (\ref{4tp12}) is
\begin{equation}\label{4tp15}
{\cal G}_{k_R}(\eta)=i\frac{\sqrt{\pi}}{2}\sqrt{-\eta}{\cal H}_{\nu _T}^{(2)}[-\eta k_{R}].
\end{equation}
This solution describes the dynamics of the effective 4D tensor modes.

\subsection{The spectrum and the scalar tensor ratio}

Since we are obtained the dynamical solution for the effective 4D tensor modes, now we are able to obtain the corresponding 
spectrum on super Hubble scales.  The amplitude of these 4D tensor metric fluctuations $<h^{2}>_{IR}$ in a de Sitter expansion on 
the IR sector $(-k_R\eta \ll 1)$ is given by
\begin{equation}\label{sp1}
\left<h^{2}(\eta)\right>_{IR}=\frac{4a^{-2}(\eta)}{\pi^{2}}\int _{0}^{\epsilon k_{H}}\frac{d 
k_{R}}{k_{R}}k_{R}^{3}\left.\left[{\cal G}_{k_R}(\eta){\cal G}_{k_R}^{*}(\eta)\right]\right|_{IR},
\end{equation}
where $\epsilon =k_{max}^{IR}/k_p\ll 1$ is a dimensionless parameter, being  $k_{max}^{IR}=k_{H}(\eta 
_{i})=\sqrt{2-m^{2}}\,\eta^{-1}_{i}$ the wave number related to the Hubble radius at the conformal time $\eta _{i}$, which 
corresponds to the time when the horizon re-enters. Here $k_p$ is the Planckian wave number. For a Hubble parameter $H=0.5\times 
10^{-9} M_{p}$ values of $\epsilon$ on the range of $10^{-5}$ to $10^{-8}$ corresponds to the number of e-foldings $N_{e}=63$ 
\cite{Gravem}.
Now we consider the asymptotic expansion for the Hankel function ${\cal H}_{\nu _T}^{(2)}[x]\simeq-(i/\pi)\Gamma (\nu  
_{T})(x/2)^{-\nu _{T}}$ in the expression (\ref{4tp15}). Thus the expression (\ref{sp1}) becomes
\begin{equation}\label{sp2}
\left<h^{2}(\eta)\right>_{IR}=\frac{2^{2\nu _T}-1}{\pi ^{3}}H^{2}\Gamma ^{2}(\nu _T)(-\eta)^{3-2\nu _T}\int _{0}^{\epsilon 
k_H}\frac{dk_R}{k_R}k_{R}^{3-2\nu _T}.
\end{equation}
We can see from this equation that the corresponding spectrum is ${\cal P}_{g}(k_R)\sim k_R^{3-2\nu _T}$, which for $m \ll 1$ is 
nearly scale invariant. In the case of tensor perturbations, the spectral index is defined by $n_{T}=3-2\nu _T$. Hence for $m \ll 
1$ the spectral index is $n_{T}\simeq 0$. Performing the remaining integration in (\ref{sp2}) we obtain
\begin{equation}\label{sp3}
\left<h^{2}(\eta)\right>_{IR}=\frac{2^{2\nu _T}-1}{\pi ^3}\frac{\Gamma^{2}(\nu _T)}{3-2\nu _T}H^{2}\epsilon ^{3-2\nu 
_T}\left(-\eta k_H\right)^{3-2\nu _T},
\end{equation}
which in terms of the cosmic time $t$ gives
\begin{equation}\label{sp4}
\left<h^{2}(t)\right>_{IR}=\frac{2^{1+2\nu _T}}{\pi}\frac{\Gamma ^{2}(\nu _T)}{3-2\nu _T}\left(\frac{H}{2\pi}\right)^{2}\epsilon 
^{3-2\nu _T}\left(\frac{k_H}{aH}\right)^{3-2\nu _T}.
\end{equation}
Furthermore, the gravitational spectrum ${\cal P}_{g}(k_{R})$ can be written in the form
\begin{equation}\label{nsp1}
{\cal P}_{g}(k_H)=\left.\frac{2^{1+2\nu _T}}{\pi}\Gamma ^{2}(\nu 
_T)\left(\frac{H}{2\pi}\right)^{2}\left(\frac{k_{R}}{aH}\right)^{3-2\nu _T}\right|_{k_H}
\end{equation}

In order to obtain the ratio of tensor to scalar modes we use the results for the spectrum of the scalar modes obtained in 
\cite{WW9}. As we have shown in \cite{WW9}, the spectral index for scalar metric fluctuations is given by 
$n_{s}=4-2\nu_{s}=4-\sqrt{9+(16k_{\psi_0}^{2}/H^{2})}$. On the other hand, we can define in analogy with the definition of 
slow-roll parameters for the 4D inflaton field, the parameter $\theta =4k_{\psi}^{2}/H^{2}<<1$ and thereby the relation $2-\nu 
_{s}\simeq \theta$ is valid \cite{ARiotto}. Hence the ratio of tensor to scalar modes $r=A_{T}^{2}(k_R)/A_{\cal R}^{2}(k_R)$ is 
\cite{WW10}
\begin{equation}\label{sp5}
r=\frac{1}{8}\theta =\frac{k_{\psi _0}^{2}}{2H^{2}}
\end{equation}
where $k_{\psi _0}$ is the wave number related with the fifth coordinate under the foliation $\psi=\psi _{0}=H^{-1}$. As we have 
shown in \cite{WW9}, the parameter $k_{\psi _0}^{2}$ is limited by the interval $0< k_{\psi _0}^{2}\leq (0.15)^{2}H^{2}$. Thus, 
the expression (\ref{sp5}) establish for $r$ the range of values $0<r\leq 0.01125$, where the value $k_{\psi 
_{0}}^{2}=(0.15)H^{2}$ corresponds to a tensor spectral index $n_{T}=-0.0225 $. On the other hand, it is known that combining the 
results of COBE with other observations on small angular scales it is obtained that $r<0.5$, at $95\%$ \cite{WW1}. Therefore, in 
this sense, our results are in agreement with observations.

\section{Final comments and conclusions}

In this letter we have studied the generation of gravitational waves during inflation from a 5D vacuum theory of gravity. 
Initiating with the 5D Riemann-flat background metric (\ref{ww1}), that describes perfectly a 5D vacuum defined by $R_{AB}=0$, we 
have studied the dynamics of the tensor fluctuations of (\ref{ww1}) introduced by the perturbed 5D line element (\ref{tp1}). In 
the non-physical coordinates $\left[N,\vec{r}(X,Y,Z),\psi\right]$ we obtained that the KK-modes satisfy the same dynamics that the 
one obtained within the braneworld formalism in \cite{WW6}. Implementing a coordinate transformation, in order to use physical 
coordinates, and considering the foliation on the fifth coordinate $\psi =H^{-1}$ (being $H$ the constant Hubble parameter) we 
obtained the effective 4D dynamics of the induced gravitational waves on an effective 4D de Sitter space-time described by 
(\ref{4tp3}). As we have shown in other works \cite{WW8,WW9}, within this formalism it is possible to geometrically induce a 4D 
effective inflationary potential $V(\varphi)=2H^{2}\varphi ^{2}$, where the deviation of the scalar spectrum from the scale 
invariance is geometrically generated due to the existence of the fifth dimension $\psi$. This fact allows to reconcile the 
constancy of the Hubble parameter $H$ with a kind of slow-roll condition introduced by the parameter $\theta$ in (\ref{sp5}), 
without the introduction by hand of a mass-term for the inflaton potential. \\

Analyzing the dynamics of the KK-modes given by (\ref{ntp9}), we obtained that on very large scales (super-Hubble scales) there is 
always a normalizable zero-mode related with a homogeneous metric perturbation on the effective 4D de Sitter background. Besides, 
as a consequence of using the 5D Riemann-flat background metric (\ref{ww1}), we have that the dynamical equation for the 5D modes 
is separable. Hence, in our case, it is possible to separate a zero-mode that corresponds to a massless graviton on the effective 
4D de Sitter background. According to (\ref{sp4}) the amplitude of gravitational waves generated during inflation is almost the 
same than the standard result obtained in typical 4D inflationary models with general relativity. Thus, we have that fluctuations 
in the massless graviton on cosmological scales are given by the Hawking temperature $H/(2\pi)$. In this sense, we recover the 
standard result obtained typically and we agree with the result of braneworld cosmology at low energies given in \cite{WW6}. 
Finally we have established the range of values for the amplitude of the tensor to scalar ratio $0<r\leq 0.01125$, which fits 
inside of the permitted range given by COBE estimations \cite{WW1}.\\

On the other hand, using the relation $n_{T}=3-\sqrt{9-4m^{2}}$, for the tensor spectral index $n_{T}=-0.0225$, it is obtained the 
square KK-mass $m^{2}=-0.033876$. According to the analysis made below equation (\ref{ntp9}), this value bears to an unstable 
KK-mode that would correspond to an unstable graviton. However, this result should be interpreted with more care. It must be 
considered the fact that the tensor spectral index $n_T$ that we have obtained using the tensor to scalar ratio (\ref{sp5}) is 
valid for the whole modes on large scale (IR sector). Thus, this $n_{T}$ has the contribution of the zero mode, that is finite and 
normalizable, and of the remaining massive KK-modes. Thereby we can interpret the value $m^{2}=-0.033876$ as a result of the 
impact of the massive KK-gravitons on the hypersurfaces $\psi =H^{-1}$. This kind of effects due to massive KK-gravitons usually 
appears in theories in higher dimensions. For instance, in the second Randall-Sundrum type brane cosmology with back-reaction 
effects included in the bulk, every KK-mode behaves as cosmic dust with a negative energy density on the brane but positive energy 
density on the bulk \cite{WW11}. A more exhaustive analysis of the behavior of the KK-gravitons within a 5D vacuum formalism with 
a noncompact fifth coordinate will be done in a forthcoming letter.

\section*{Acknowledgements}
\noindent
JEMA acknowledges CNPq-CLAF and UFPB (Brazil) for financial
support. \\

\end{document}